\documentclass[12pt]{article}
\usepackage{graphicx}
\usepackage{latexsym}
\usepackage{amscd}
\usepackage{amsmath}
\usepackage{amssymb}
\usepackage[small,centerlast]{caption2}
\usepackage{longtable}
\usepackage[dvipsnames]{color}% - это необходимо для правки
\usepackage{cite}

\graphicspath{{Figures/}}
\def\red#1{\textcolor{red}{#1}}

\makeatletter
\renewcommand{\section}{\@startsection {section}{1}{\z@}%
	{-3.5ex \@plus -1ex \@minus -.2ex}%
	{2.3ex \@plus.2ex}%
	{\normalfont\Large}}
\renewcommand{\subsection}{\@startsection{subsection}{2}{\z@}%
	{-3.25ex\@plus -1ex \@minus -.2ex}%
	{1.5ex \@plus .2ex}%
	{\normalfont\large\itshape}}
\renewcommand{\subsubsection}{\@startsection{subsubsection}{3}{1em}%
	{-3.25ex\@plus -1ex \@minus -.2ex}%
	{-1.5em \@plus .2em}%
	{\normalfont\normalsize\bfseries}}

\makeatother

\begin{document}
\begin{center}
	
	{\Large \textbf{Electromagnetic and Centrifugal Effects on Plasma Acceleration in the Magnetic Nozzle\\}}
	
	\medskip
	
	{\large \textbf{A.I. Smolyakov$^{a,*}$, A. Sabo$^a$, S.I.~Krasheninnikov$^b$, P.N.~Yushmanov$^c$}}
	
	\medskip
	
	\textit{$^a$\,University of Saskatchewan, Saskatoon, Canada}\\
	\textit{$^b$\,University of California San Diego,  California, USA}\\
	\textit{$^c$\,TAE Technologies Inc., Foothill Ranch, California, USA}\\
	\textit{$^*$\,e-mail: andrei.smolyakov@usask.ca}
	\\[10pt]
	%{\footnotesize Поступила в редакцию 03.07.2009~г.$\;$\\
	%	окончательный вариант получен 04.03.2009~г. }
\end{center}

\renewcommand{\abstractname}{}
\begin{abstract}
	\noindent\small Plasma flow and acceleration in the converging-diverging magnetic field configuration, such as magnetic nozzle in 
	electric propulsion and open magnetic mirrors for fusion applications are considered. This work analyses plasma acceleration in the magnetic nozzle with an emphasis on the electromagnetic  effects and  centrifugal forces due to plasma rotation. Intrinsic coupling of the azimuthal rotation and azimuthal magnetic field is analyzed, and additional plasma acceleration due to the conversion of the energy of the azimuthal magnetic field and azimuthal rotation is demonstrated. For large expansion in the diverging magnetic field plasma flow velocities may approach and exceed the Alfven velocity. In these regimes, stationary solutions for the transonic and trans-Alfvenic flows have been obtained that demonstrate the existence of the unique regular solution passing through all critical points within the MHD theory, i.\,e.  the points  where the plasma flow is equal to the signal velocities of the MHD modes: slow and fast magnetohydrodynamic waves and Alfven wave. The time-dependent initial value simulations show that stationary equilibrium flows are robust and stable, so that time-dependent solutions converge toward stationary solutions.    
	
\end{abstract}

\section{Introduction}
Converging-diverging magnetic nozzle configurations are employed in plasma propulsion systems [1--4] and also in open mirror systems for fusion applications [5, 6]. For propulsion applications, one function of the magnetic nozzle is similar to the operation of gas-dynamics Laval nozzle where the nozzle converts the energy of the thermal (random) kinetic energy into the kinetic energy of the directed axial flow
producing the thrust. In fusion systems, the mirror magnetic field configurations are used to confine plasmas, while the diverging part (the divertor or expander) allows to reduce the energy load per unit of the wall area [5, 6]. In both propulsion and fusion oriented  applications, the plasma is accelerated transonically in the mirror. A detailed understanding of the mechanisms and physics of plasma flow and acceleration in the mirror configuration is important for the optimization of plasma thrust in propulsion and the determination  of the axial plasma and energy losses in the open mirror confinement systems.     
This work focuses on the additional roles of the magnetic nozzle in converting the energy of azimuthal plasma rotation and the magnetic energy associated with the self-consistent plasma current into the kinetic energy of the axial flow— the  process known as swirl and electromagnetic acceleration \red{[7]}.   
In the presence of the longitudinal magnetic field, the azimuthal plasma rotation and azimuthal magnetic field represent essential polarization of the circular polarized Alfven wave. The finite plasma pressure and compressibility couple them to the perturbations of the magnetosonic type. Overall, the magnetic nozzle physics includes the self-consistent 
coupling of the plasma pressure, axial plasma velocity, azimuthal rotation, and azimuthal magnetic field due to the axial current. This physics thus represents the dynamics of Alfven, fast, and slow magnetosonic waves in flowing plasma in the inhomogeneous magnetic field, which can be analyzed in the framework of one-fluid ideal Magnetohydrodynamic (MHD) model \red{[1, 2, 8--10]}.  

In general, the problem is two-dimensional and involves the solution of nonlinear differential equations generalizing the Grad--Shafranov equation including inertial forces [1, 9, 11--13].  Here, we present a quasi-two-dimensional (paraxial) model for the coupled
evolution of plasma pressure, axial velocity $V_{\Vert }$, azimuthal rotation $V_{\phi }$, and
toroidal (azimuthal) magnetic field $B_{\phi }$ similar to [1, 2, 8--10]. 
The model includes the effects of plasma acceleration by the pressure and conversion of the energy of the azimuthal magnetic field and rotation into the energy of the directed axial motion.  
The time-dependent
model is used to study the effects of the current and rotation
propagation from the point of the power input (by the rotation and Poynting
flux). In the stationary case, the equations can be integrated resulting in a system of
algebraic equations for $V_{\Vert }$, $V_{\phi }$, $B_{\phi }$ and $p$.\
Solutions of these equations give the profiles $V_{\Vert }$, $V_{\phi }$, $%
B_{\phi }$ and $p$ as functions of the poloidal magnetic field $B_{p}$
(equivalently as a function of the axial coordinate). In this paper, we analyze the structure of the steady-solutions, presenting a full diagram of stationary solutions resolving the critical points, investigate how the rotation and azimuthal magnetic field affect the acceleration, and study the stability of stationary solutions by time-dependent analysis.

\section{Basic equations and quasi-two-dimensional (paraxial) model}

We use standard one-fluid MHD equations: the continuity equation
\begin{equation}
\frac{{\partial \rho }}{{\partial t}}+\nabla \cdot \left( {\rho }\mathbf{V}%
\right) =0,
\end{equation}%
the plasma momentum balance
\begin{equation}
\rho \left( \frac{{\partial }\mathbf{V}}{{\partial t}}+\left( \mathbf{V}%
\cdot {\nabla }\right) \mathbf{V}\right) =\frac{1}{c}\mathbf{J\times B}%
-\nabla p=-\frac{1}{8\pi }\nabla B^{2}+\frac{1}{4\pi }\left( \mathbf{B\cdot }%
\nabla \right) \mathbf{B}-\nabla p,
\label{mb}
\end{equation}%
\ and the ideal Ohm's law
\begin{equation}
\left( \mathbf{E}+\frac{1}{c}\mathbf{V}{\times }\mathbf{B}\right) =0,
\label{ohm}
\end{equation}%
written in the form of frozen-in condition for the magnetic field
\begin{equation}
\frac{{\partial }\mathbf{B}}{{\partial t}}=\nabla \times \left( \mathbf{V}{%
\times }\mathbf{B}\right) .  \label{ind}
\end{equation}

The system has to be closed with the pressure/equation of state model. In the ideal MHD, the
plasma pressure is a sum of the electron and ion pressure, which have rather
different behavior. Generally, in such regimes local (fluid type) closures
for heat fluxes do not exist and/or are not reliable. As a simplification,
two opposite extreme situations, adiabatic and isothermal, are often
considered. Both species are expected to be anisotropic in the weakly
collisional regimes of interest. In the adiabatic limit, assuming that heat
fluxes can be neglected for ions, but taking into account the pressure
anisotropy, one arrives at the two-pressure adiabatic Chew--Goldberger--Law
model. The double adiabatic pressure model was used to describe the effects
of the anisotropic ion temperature on plasma acceleration in the magnetic
nozzle in [14].

 Due to their high mobility, electrons are expected to have large heat fluxes.
The isothermal limit is justified for an infinite heat conductivity
when the temperature formally remains uniform to avoid the diverging energy flux.
 In the conditions of the magnetic nozzle, plasma acceleration and expansions results in
the density gradient supported by the electric field which reflects a large
fraction of electrons. Therefore, a Boltzmann distribution with uniform
temperature is often used as a simplification for electrons. In reality the electron pressure is anisotropic in
the strong magnetic field, and the conversion of the electron thermal (random) energy into the kinetic energy of the accelerated plasma must result in a decrease of the electron temperature. An additional cooling of the electrons will occur due to radial losses and axial heat fluxes. 
Experimental measurements report electron cooling with an effective
adiabatic exponent [15, 16]  $\gamma >1$.  For simplicity, one  can make an
assumption of the isotropic pressure with a general polytropic
equation of state
\begin{equation}
\frac{{\partial p}}{{\partial t}}+\nabla \cdot \left( {p}\mathbf{V}\right)
+\left( \gamma -1\right) p\nabla \cdot \mathbf{V}=0,  \label{en}
\end{equation}%
where $\gamma $ is some constant between $\gamma =5/3$ (adiabatic) and $%
\gamma =1$ (isothermal) cases.

The nozzle magnetic field is defined in the cylindrical coordinate system, $%
\mathbf{B}=\left( {{B_{r}},{B_{\phi }},{B_{z}}}\right) $, which can be
written as

\begin{equation}
\mathbf{B}={B_{p}}\mathbf{b}+{B_{\phi }}\widehat{\mathbf{e}}_{\phi },  \label{b}
\end{equation}%
where ${B_{p}}={({B_{z}}^{2}+{B_{r}}^{2})^{1/2}}$ is the field in the radial-axial
(poloidal) plane, and $\mathbf{b}=\left( {{B_{z}}\widehat{\mathbf{e}}{_{z}}+{B_{r}}%
\widehat{\mathbf{e}}{_{r}}}\right) /\;{B_{p}}$ is the corresponding unit vector in the
$\left( r,z\right) $ plane. One can show that within the one-fluid MHD theory, the plasma flows along the magnetic flux surfaces [11], so that $\mathbf{V}_{p}$ and $\mathbf{B}_{p}$ are parallel to each, $\mathbf{V}_{p}\Vert $ $\mathbf{B}_{p}$.  Then, plasma velocity can also be represented in the form
\begin{equation}
\mathbf{V}={V_{p}}\mathbf{b}+{V_{\phi }}\widehat{\mathbf{e}}{_{\phi },}  \label{v}
\end{equation}%
${V_{p}}={(}V{{_{z}}^{2}+}V{{_{r}}^{2})^{1/2}}$. A main assumption is that the flow
is symmetric in the $\phi $, azimuthal (toroidal) direction, $%
\partial /\partial \phi =0$.  In general, when
the Hall term and electron pressure are added to Ohm's law, plasma flow
velocity $\mathbf{V}_{p}$ and magnetic field $\mathbf{B}_{p}$ are not
parallel. 

With equations (6) and (7), the continuity equation takes the form
\begin{equation}
\frac{\partial }{{\partial t}}{\rho }+B_{\Vert }{\nabla _{\parallel }}\left(\frac{%
{\rho V_{\parallel }}}{B_{\Vert }}\right)=0,
\end{equation}%
where we use the notation $B_{p}\equiv B_{\Vert }$, ${V_{p}\equiv
V_{\parallel }}$, and $\nabla_\Vert= \mathbf{B}_{p}\cdot \nabla /B_p$.  Projecting the momentum balance equation (\ref{mb}) and
the induction equation (\ref{ind}) on the toroidal direction $\widehat{%
\mathbf{e}}_{\phi }$ one obtains
\begin{equation}
\frac{\partial }{{\partial }t}V_{\phi }+V_{\parallel }\nabla _{\parallel
}V_{\phi }+\frac{{V}_{r}V{{_{\phi }}}}{r}=\frac{B{{_{\parallel }}}}{{4\pi
\rho }}{\nabla _{\parallel }}B_{\phi }+\frac{B{{_{r}}}B{{_{\phi }}}}{{4}\pi
\rho r}\;,
\end{equation}%
\begin{equation}
\frac{\partial }{{\partial t}}B_{\phi }+V_{\parallel }\nabla _{\parallel
}B_{\phi }+\frac{V_{\phi }{B}_{r}}{r}=B_{\parallel }\nabla _{\parallel
}V_{\phi }+\frac{{{\ B_{\phi }V}}_{r}}{{r}}-B_{\phi }\left( V_{\Vert }\nabla
\cdot \mathbf{b+}\nabla _{\parallel }V_{\Vert }\right) .  \label{bf0}
\end{equation}

Projecting the momentum balance equation on the poloidal plane gives the equation for the poloidal flow velocity driven by
the pressure gradient and ponderomotive force from the azimuthal (toroidal)
magnetic field 
\begin{equation}
\frac{\partial }{{\partial }t}V_{\parallel }+V_{\parallel }\nabla
_{\parallel }V_{\parallel }-\frac{B{{_{\mathbf{r}}}}}{rB{{_{\parallel }}}}%
V_{\phi }^{2}=-\frac{1}{{4\pi \rho }}\frac{B_{\mathbf{r}}}{rB{{_{\parallel }}%
}}B_{\phi }^{2}-\frac{1}{4\pi \rho }\nabla _{\parallel }B_{\phi }^{2}-\frac{1%
}{\rho }\nabla _{\parallel }p.
\end{equation}%
Note that due to the inhomogeneous magnetic field, there are additional
contributions to the poloidal projections from the centrifugal force (the last
term on the left-hand side) and the ponderomotive magnetic field force (the first term on the right-hand side).

We note that equations (8)--(11) are exact two-dimensional equations, including the two-dimensional differential operator $\nabla_\Vert$ in the poloidal plane $r,z$. 
These equations can be reduced in the paraxial approximation, where the two-dimensional effects are included to first-order terms [1, 2] in the parameter $r/a<1$, where $a$ is the characteristic
radial dimension. This reduction is done in two steps. We note that in the first order, the expressions for the radial magnetic field and radial plasma velocity can be approximated as 
\begin{equation}
\frac{1}{r}\frac{\partial }{\partial r}rB_{r}+\frac{\partial }{\partial z}%
B_{z}=0,
\end{equation}%
\begin{equation}
{B_{r}}\simeq -\frac{r}{2}\frac{{\partial B}_{z}\left( z\right) }{{\partial z%
}},
\end{equation}%
and respectively,
\begin{equation}
{V_{r}}={V_{\parallel }}\frac{{{B_{r}}}}{{{B_{\parallel }}}}=-V_{\Vert }%
\frac{r}{2B_{\Vert }}\frac{{\partial B}}{{\partial z}},
\end{equation} where $V_\Vert$ and $B_\Vert$ are the total plasma velocity and magnetic field in the magnetic surface, as defined in Eq. (8).
 Using these approximations, and $\nabla \cdot \mathbf{b}=-\nabla _{\Vert
}B_{\Vert }/B_{\Vert }$, 
equations (8)--(11) take the form  
\begin{equation}
\frac{\partial }{{\partial t}}{\rho }+B_{\Vert }{\nabla _{\parallel }}(\frac{%
{\rho V_{\parallel }}}{B_{\Vert }})=0,  \label{vc}
\end{equation}

\begin{equation}
\frac{\partial }{{\partial }t}V_{\phi }+V_{\parallel }\nabla _{\parallel
}V_{\phi }-\frac{B{{_{\parallel }}}}{{4\pi \rho }}{\nabla _{\parallel }}%
B_{\phi }=\frac{1}{2}\left( V_{\parallel }V{{_{\phi }}}-\frac{B{{_{\phi }}%
B_{\parallel }}}{{4}\pi \rho }\right) \frac{\nabla _{\parallel }B_{\Vert }}{%
B_{\Vert }}\;,  \label{vf2}
\end{equation}%
\begin{equation}
\frac{\partial }{{\partial t}}B_{\phi }+V_{\parallel }\nabla _{\parallel
}B_{\phi }+B_{\phi }\nabla _{\parallel }V_{\Vert }-B_{\parallel }\nabla
_{\parallel }V_{\phi }=\frac{1}{2}\left( B_{\phi }V_{\Vert }+V_{\phi
}B_{\Vert }\right) \ \frac{\nabla _{\parallel }B_{\Vert }}{B_{\Vert }}.
\label{bf2}
\end{equation}%
\begin{equation}
\frac{\partial }{{\partial }t}V_{\parallel }+V_{\parallel }\nabla
_{\parallel }V_{\parallel }+\frac{1}{2\ }V_{\phi }^{2}\frac{\nabla
_{\parallel }B_{\Vert }}{B_{\Vert }}=-\frac{1}{\rho }\nabla _{\parallel }p+%
\frac{1}{{8\pi \rho }}\frac{\nabla _{\parallel }B_{\Vert }}{B_{\Vert }}%
B_{\phi }^{2}-\frac{1}{8\pi \rho }\nabla _{\parallel }B_{\phi }^{2}.
\label{vp}
\end{equation}
Technically, in these equations, the $\nabla _\Vert$ operator is still two-dimensional and represents the gradient along the magnetic surface in the poloidal plane. All plasma parameters, such as $V_\Vert, p, V_\phi$, and $ B_\phi $  are defined in the two-dimensional space ($r-z$) along the magnetic flux surfaces. In the paraxial approximation, the magnetic field $\mathbf{B}_p=r^{-1} \nabla \psi\times  {\bf e}_\phi$, is described by the magnetic flux function in the form $\psi(r,z)=r^2 B(z)/2$. Then, near the axis, one has $B_\Vert \simeq B(z)$, and the length along the magnetic surface can be approximated by the $z$-coordinate, $\nabla _\Vert \simeq \partial/\partial z$,  thus equations (15), (16) become one-dimensional and can be resolved in the $z$-space. It is worth to emphasize however that these equations retain variations of the plasma parameters in the $(r,z)$ space that occur along the two-dimensional magnetic surfaces $\psi=\psi(r,z)$.  
The radial dependence of plasma parameters, in particular, azimuthal rotation and magnetic field is further discussed in Section 7.3.  

Together with the equation of state, equations (15)--(18) describe the coupled nonlinear evolution of plasma acceleration $V_\Vert$, plasma pressure $p$, plasma rotation $V_\phi$, and azimuthal magnetic field  $B_\phi$. We note that in this model, the poloidal magnetic field $B_\Vert (z)$ does not change and is assumed as given.  

\section{Alfven waves in a homogeneous plasma with flow}

It is instructive to highlight the Alfven wave physics by considering the limit case of
the uniform magnetic field $B{{_{\parallel }}}$. Then with $\nabla
_{\parallel }B_{\Vert }\rightarrow 0$ and $V_{\parallel }\rightarrow 0,$ one has
\begin{equation}
\frac{\partial }{{\partial }t}V_{\phi }=\frac{B{{_{\parallel }}}}{{4\pi \rho
}}{\nabla _{\parallel }}B_{\phi }
\end{equation}%
and%
\begin{equation}
\frac{\partial }{{\partial t}}B_{\phi }=B_{\parallel }\nabla _{\parallel
}V_{\phi }.\
\end{equation}

These equations describe the evolution of azimuthal plasma rotation $V_{\phi
}$ and magnetic field $B_{\phi }$ in the circularly polarized Alfven wave.
With a finite $V_{\parallel}$ and retaining nonlinear terms, one obtains
\begin{equation}
\frac{\partial }{{\partial }t}V_{\parallel }+V_{\parallel }\nabla
_{\parallel }V_{\parallel }+\frac{1}{8\pi \rho }\nabla _{\parallel }B_{\phi
}^{2}=0,
\label{vp}
\end{equation}%
\begin{equation}
\frac{\partial }{{\partial }t}V_{\phi }+V_{\parallel }\nabla _{\parallel
}V_{\phi }-\frac{B{{_{\parallel }}}}{{4\pi \rho }}{\nabla _{\parallel }}%
B_{\phi }=\ 0,
\label{vphi0}
\end{equation}%
\begin{equation}
\frac{\partial }{{\partial t}}B_{\phi }+V_{\parallel }\nabla _{\parallel
}B_{\phi }+B_{\phi }\nabla _{\parallel }V_{\Vert }-B_{\parallel }\nabla
_{\parallel }V_{\phi }=\ 0.
\end{equation}%
These nonlinear equations describe plasma acceleration by the ponderomotive
force of the Alfven wave, $\nabla _{\parallel }B_{\phi }^{2}/8\pi \rho $.
Alternatively, one can consider pumping the Alfven waves (the azimuthal field
$B_{\phi })$ by the plasma flow along the inhomogeneous magnetic field. The effects of the inhomogeneous magnetic field in the general case bring additional and nontrivial coupling with plasma rotation and azimuthal magnetic field as it is shown in Eqs. (16)--(18). We note that stationary solutions for the plasma flow considered below retain the essential physics of the Alfven waves manifested  by the coupling  of the azimuthal rotation and azimuthal magnetic field.  

\section{Energy conservation and Bernoulli integral}

It is useful to highlight the relation of the energy conservation  to one of the of conserved quantity--- the Bernoulli integral consider below.  The thermal energy conservation is written in Eq. (\ref{en}). Summing
it with the magnetic field energy from (\ref{ind}) and the kinetic energy from (\ref{vp}) and (\ref{vphi0}),  one obtains the total
energy conservation equation 
\begin{equation}
\frac{{\partial }}{{\partial t}}\left( \rho \mathbf{V}^{2}/2+\frac{\mathbf{B}%
^{2}}{8\pi }+\frac{p}{\gamma -1}\right) +\nabla \cdot \left( \mathbf{V}%
\left( \rho \frac{\mathbf{V}^{2}}{2}+\frac{\gamma p}{\gamma -1}\right) +%
\frac{1}{4\pi }\mathbf{E\times B}\right) =0.
\end{equation}%
The Poynting energy flux can be rewritten with (\ref{ohm}) giving
\begin{equation}
\frac{{\partial }}{{\partial t}}\left( \rho \mathbf{V}^{2}/2+\frac{\mathbf{B}%
^{2}}{8\pi }+\frac{p}{\gamma -1}\right) +\nabla \cdot \left( \mathbf{V}%
\left( \rho \frac{\mathbf{V}^{2}}{2}+\frac{B^{2}}{4\pi }+\frac{\gamma p}{%
\gamma -1}\right) -\frac{1}{4\pi }\mathbf{B}\left( \mathbf{V\cdot B}\right)
\right) =0.
\end{equation}%
This general equation in two-dimensional form can be reduced to one-dimensional form by explicitly taking into account that the plasma flows along the magnetic surfaces.  After some algebra, using Eqs. (6). (7),  in stationary state $\partial /\partial t=0,$ one 
obtains the following conserved integral representing the constant energy flux along the magnetic flux surface 
\begin{equation}
I_{B}=\frac{V_{\Vert }}{B_{\Vert }}\left( \rho \frac{V_{\Vert }^{2}}{2}+\rho
\frac{V_{\phi}^{2}}{2}+\frac{B_{\phi }^{2}}{4\pi }+\frac{\gamma p}{\gamma -1%
}\right) -\frac{1}{4\pi }V_{\phi }B_{\phi }.
\end{equation}%
Thus, additional terms due to the azimuthal magnetic field and rotation are related to the Poynting energy flux. In the paraxial limit, the conserved quantity can be rewritten in the form of the Bernoulli
integral given in the next section.
\section{Summary of stationary equations}
As noted above, near the axis, the distance along the curved magnetic surface can be approximated by the $z$-coordinate. Then, in the stationary state $\partial /\partial t=0$, equations (\ref{vc})--(\ref{vp}) can be integrated in $z$ giving
\begin{equation}
\frac{\rho V_{\Vert }}{B_{\Vert }}=\Gamma ,  \label{s1}
\end{equation}%
\begin{equation}
{V_{\parallel }B_{\phi }-V_{\phi }B_{\parallel }=}\sqrt{B_{\Vert }}\beta,
\label{beta}
\end{equation}%
\begin{equation}
\frac{\Gamma V_{\phi }-B_{\phi }/4\pi }{\sqrt{B_{\Vert }}}=W,  \label{W}
\end{equation}
\begin{equation}
I=\ \ \frac{V_{\Vert }^{2}}{2}+\frac{V_{\phi }^{2}}{2}+\frac{B_{\phi }^{2}}{%
4\pi \rho }+\ \frac{\gamma }{\gamma -1}\frac{p}{\rho }-\frac{1}{4\pi \Gamma }%
V_{\phi }B_{\phi }.
\end{equation}%
 These equations are supplemented by
an additional conservation law for the effective polytrope
\begin{equation}
S=\frac{p}{\rho ^{\gamma }}.  \label{sg}
\end{equation}

We recall that these equations describe the magnetic nozzle in the paraxial
approximation in which the the plasma parameters vary along the two-dimensional flux surfaces of the magnetic field  tube with a given value of the magnetic flux, as shown in Fig. 1. The radial dependence in these equations is implicit in the integration constants and inverse function $B_\Vert=B_\Vert (\psi,r)$.

Equation (\ref{s1}) describes the conservation of matter for the plasma flow inside the magnetic flux tube with varying cross section $S$. Note that due to the magnetic flux conservation, $BS=\text{const}$. 
Equation (\ref{beta})  follows from (\ref{bf2}), it is a consequence of the azimuthal symmetry, and the constant $\beta$ in this equation is related to the radial electric field. Equation (\ref{W}) follows from (\ref{vf2}) and is the generalization of the conservation of the angular momentum in presence of the poloidal flow $V_\Vert$. Many authors have used the term ``specific angular momentum'' for the conserved quantity $W$. However, it has been suggested  recently \red{[17]} to abandon this misnomer to avoid the confusion with the specific angular momentum in the relativistic case. 
Equation (30) is a consequence of the energy conservation (26) and Eq. (27).
 These equations can also be derived
from general two-dimensional $(r,z)$ equations for
axisymmetric plasma flows \red{[11-13]}. The angular integrals of motion (\ref{beta}) and (\ref{W}) are similar to those obtained in the theory of magnetic winds \red{[18]}, also see Refs. [1, 2, 8--10].

\section{Critical points and Alfven wave singularity}

Algebraic equations (\ref{s1})--(\ref{sg}) fully define the axial profiles of
plasma variables $V_{\Vert }$, $V_{\phi }$, $B_{\phi }$ and $p$. Boundary
conditions at the breach of the magnetic nozzle determine the values of the
integrals $\Gamma ,W,\beta$, $I$, and $S$ which in turn determine the solutions
for\ $V_{\Vert }(z)$, $V_{\phi }(z)$, $B_{\phi }\left( z\right) $ and $%
p\left( z\right) $ for a given function of $B_{\Vert }(z).$ There exist multiple  solutions in the ``phase''
space diagram $(V_\Vert(z),z)$. This diagram has three critical points at which $\partial
V_{\Vert }/\partial z$ may diverge. One can show [1, 2] that two critical points are
determined by the condition
\begin{equation}
D=\left( V_{\Vert }^{4}-V_{\Vert }^{2}\left( c_{s}^{2}+c_{A}^{2}\right)
+c_{s}^{2}c_{A\Vert }^{2}\right) =0,
\label{D}
\end{equation}%
where
\begin{equation}
c_{A}^{2}=c_{A\phi }^{2}+c_{A\Vert }^{2},
\end{equation}%
and
\begin{equation}
c_{A\phi }^{2}=\frac{B_{\phi }^{2}}{4\pi \rho },
\end{equation}%
\begin{equation}
c_{A\Vert }^{2}=\frac{B_{\Vert }^{2}}{4\pi \rho },
\end{equation}%
\begin{equation}
c_{s}^{2}=\frac{\gamma p}{\rho }.
\end{equation}%
These critical points correspond to the slow and fast magnetosonic ``signal''
points, i. e. points where the local plasma velocity $V_{\Vert }$ is equal to
the phase velocity of slow or fast magnetosonic mode, $c_{sm}$ and $c_{fm}$ found as the roots of equation (\ref{D}). 

An additional critical point of a different type exists at the Alfven
point where the local plasma velocity $V_{\Vert }$ is equal to the Alfven
velocity $V_{\Vert }=c_{A}$. Specifically, the condition for Alfven critical
point is $\Gamma V_{\Vert }-B_{\Vert }/4\pi =0$, which follows from the
equations for azimuthal magnetic field and azimuthal rotation
\begin{equation}
B_{\phi }=\ \sqrt{B_{\Vert }}\frac{\beta \Gamma  +B_{\Vert }W}{\Gamma
V_{\Vert }-B_{\Vert }/4\pi }\ ,
\label{bphi}
\end{equation}%
\begin{equation}
V_{\phi }=\ \sqrt{B_{\Vert }}\frac{W V_{\Vert }+\beta/4\pi }{\Gamma V_{\Vert
}-B_{\Vert }/4\pi }.
\label{vphi}
\end{equation}

% \begin{figure}[htp]
%     \centering
% \includegraphics[width=0.4\textwidth]{D4.eps}
%  \label{fig:sidebyside}
% \end{figure}

As was discussed in Ref. \red{[19]}, in the electrostatic limit and $B_{\phi }=V_{\phi
}=0$, the regularity condition at the sonic point $V_{\Vert }=c_{s},$ which
occurs at $\partial B_{\Vert }/\partial z=0,$ defines a unique regular
transonic solution for $V_{\Vert }=V_{\Vert }\left( z\right) $ for a given
magnetic field profile $B_{\Vert }\left( z\right) $.
When electromagnetic effects are included, and for a low plasma pressure, the former sonic point  $V_{\Vert }=c_{s}$ becomes the slow magnetosonic point, $V_\Vert=c_{sm}$.  An additional critical point occurs at the fast-magnetosonic signal condition $V_\Vert=c_{fm}$. For low plasma pressure, the fast magnetosonic critical point is close to the Alfven point. A unique transonic and trans-Alfvenic solution exists that is regular at all three points.

An example of a ``phase-space'' diagram for the multiple branches of $V_\Vert (z) $ solutions is shown in Fig.~2. For simplicity, for converging-diverging parts of the nozzle, here, we use a model magnetic field profiles in the form
\begin{eqnarray}
B_{\Vert }\left( z\right)  &=&\frac{B_{m}}{1+\left( R-1\right)
(z/z_{0}-1)^{2}},\text{ for }0<z<z_{0}, \\
B_{\Vert }\left( z\right)  &=&\frac{B_{m}}{1+\left( K-1\right)
(z/z_{0}-1)^{2}},\text{ for }z>z_{0}.
\label{Bm}
\end{eqnarray}

The normalized values of integrals  used in Fig. 2 are  
\begin{equation}
    \Gamma' = \frac{B_{\parallel 0}}{m_{i}n_{0}c_{s}}\Gamma= 0.0824, 
    \label{Gamma_normalized}
\end{equation}
\begin{equation}
     \beta' = \frac{\beta}{\sqrt{B_{\parallel 0}}c_{s}}=0.5631,
\end{equation}
\label{beta_normalized}
\begin{equation}
     W' = \frac{B^{3/2}_{\parallel 0}}{m_{i}n_{0}c^2_{s}}W=-3.1373.
     \label{W_normalized}
\end{equation}

 The electron pressure term in the Bernoulli integral Eq. (30) was simplified in the limit $\gamma \rightarrow 1$
\begin{equation}
    \frac{\gamma}{\gamma - 1}\frac{p_{e}}{\rho} \rightarrow c^2_{s}\ln\left(\frac{n}{n_{0}}\right) = c^2_{s}\ln\left(\frac{{B'_{\parallel}}\Gamma'}{M_{\parallel}}\right).
    \label{electron_pressure_term}
\end{equation}
The modifications for a general polytropic constant $\gamma \neq 1$ were studied in Ref. [19], and they are not significant for the purposes of this work. Plasma density and dimensional plasma pressure (beta parameter) at $z/L = 0$ were $n_{0} = 1.0 \times 10^{19}$ m$^{-3}$ and $\beta_{p} = {c^2_{s}}/{c^2_{A\parallel 0}} = 0.04$, respectively. 
 A unique transonic and trans-Alfvenic solution (shown in red in Fig. 1), which passes smoothly through all three critical points, corresponds to the unique value of the Bernoulli integral, $I^\prime =I/c_s^2=21.472$.  

\section{ Examples of stationary solutions}
In this section we consider several examples that illustrate plasma acceleration in the magnetic nozzle, including the direct effect of plasma rotation and azimuthal magnetic field. We note that these examples are also relevant to open mirror systems with biasing and neutral beam injection [6, 20], where the azimuthal rotation at the breach of the nozzle can be induced by the beams and radial electric field, and the azimuthal magnetic field is a result of the axial bias current.

\subsection{Analytical solution for cold ions and isothermal electrons}

    To establish a simplest base case model, one can consider the acceleration of
cold ions in the assumption of the isothermal electrons with $B_{\phi
}=V_{\phi }=0$, $p=T_{e}n,$ and $c_{s}^{2}=T_{e}/m_{i}.$ Then, one can
obtain for the plasma velocity [19]
\begin{equation}
M\left( z\right) =V_{\Vert }/c_{s}=\left[ -W(-b^{2}\left( z\right)) /e\right]
^{1/2},  \label{m}
\end{equation}%
where $b\left( z\right) $ is the relative magnitude of the magnetic field
normalized to the maximum value $B_{z}=B_{m}$ at $z=z_{m},$
\begin{equation}
b\left( z\right) =B\left( z\right) /B_{m}.
\end{equation}%
Here $W\left( y\right) $ is the Lambert function defined by the equation $W\exp
\left( W\right) =y.$ In the real space, the Lambert function has two branches
$W\left( 0,y\right) $ and $W\left( -1,y\right) $ that smoothly match at $%
y=1/e,$ where $e$ is Euler's constant. Two branches match at the sonic point $%
z=z_{m}$ where $B\left( z_{m}\right) =B_{m}$ and $\partial B/\partial z=0$. Other properties and interesting applications of Lambert functions in the physics of plasmas can be found in Refs. [21, 22]. 

We recall that we consider the converging-diverging magnetic nozzle
configuration. The magnetic field at the nozzle breach, $z=0,$ is denoted as
$B_{0}=B\left( 0\right) $, and the magnetic mirror ratio $R=B_{m}/B_{0}$.
The expansion ratio is defined as $K=B_{m}/B\left( L\right) $, where $%
B\left( L\right) $ is the magnetic field at the nozzle exit, $z=L,$ $\ $\
where $L$ is the nozzle length. Respectively, the transonic acceleration
profile is uniquely defined \ by equation (\ref{m}), so that the accelerating velocity profile has no free parameters, and the values of
the Mach numbers at the nozzle breach,
\begin{equation}
M\left( 0\right) =\left[ -W(0,-R^{-2}/e\right] ^{1/2},
\end{equation}%
and
\begin{equation}
M\left( L\right) =\left[ -W(-1,-K^{-2}/e\right] ^{1/2},
\end{equation}%
at the nozzle exit are unique.

\subsection{Swirl acceleration in the isothermal Laval nozzle}

To highlight the physics of the effects of centrifugal pressure due to rotation (swirl acceleration)  we consider it within the simple
two-fluid model with cold ions, isothermal electrons and assuming no electromagnetic effects such that $B_{\phi }\equiv 0.$

Plasma (ion) acceleration along the magnetic field is determined by the
projection of the ion momentum balance on the magnetic field direction
\begin{equation}
m_{i}\mathbf{b}\cdot \left( \frac{{\partial }\mathbf{V}}{{\partial t}}%
+\left( \mathbf{V}\cdot {\nabla }\right) \mathbf{V}\right) =\mathbf{b}\cdot
\mathbf{E}.  \label{mbp}
\end{equation}
Taking into the account that the electric field is balanced by the electron
pressure gradient due to expanding plasma one has
\begin{equation}
V_{\parallel }\nabla _{\parallel }V_{\parallel }=-c_{s}^{2}\nabla
_{\parallel }\ln n-\frac{1}{2\ }V_{\phi }^{2}\frac{\nabla _{\parallel }B}{B}.
\label{1}
\end{equation}%
The last term on the right-hand side is the effect of the centrifugal
pressure due to the azimuthal rotation $V_{\phi }.$ The equation for the
azimuthal rotation takes the form

\begin{equation}
V_{\parallel }\nabla _{\parallel }V_{\phi }-\frac{1}{2}V_{\parallel }V{{%
_{\phi }}}\frac{\nabla _{\parallel }B}{B}=0.  \label{2}
\end{equation}%
Equation (\ref{2}) for the azimuthal rotation can be integrated, resulting in
the conservation of angular momentum in the form%
\begin{equation}
\frac{V_{\phi }^{2}}{2B}=\mu =\text{const}.
\end{equation}%
This relation reduces to the momentum conservation $V_{\phi }r=\text{const}$ when
the magnetic flux conservation condition $Br^{2}=\text{const}$ for the magnetic
tube in the paraxial model is used.

Equations (\ref{1}) can be reduced to the form

\begin{equation}
\left( V_{\parallel }-\frac{c_{s}^{2}}{V_{\parallel }}\right) V_{\Vert
}^{^{\prime }}=-\left( c_{s}^{2}+\frac{1}{2\ }V_{\phi }^{2}\right) \frac{%
\nabla _{\parallel }B}{B},  \label{vp0}
\end{equation}%
where $\partial \left( ...\right) /\partial z=\left( ...\right) ^{^{\prime
}} $. This equation demonstrates that finite azimuthal rotation $%
V_{\phi }$ increases plasma acceleration.

Equation (\ref{vp0}) shows that the energy of the azimuthal rotation is a
positive additive to the electron thermal energy. Similar to the Laval
nozzle effect, the converging-diverging magnetic mirror configuration converts the plasma thermal energy and the energy of the ion azimuthal
rotation into the kinetic energy of the axial ion motion, $V_{\Vert }^{2}/2.$

We note here that, as it could be expected, the role of the azimuthal
rotation in Eq. (\ref{vp0}) is analogous to the role of a finite
perpendicular ion pressure. As it was shown in Ref. \red{[14]},
finite perpendicular pressure $p_{\bot }$ increases the ion acceleration
similarly to Eq. (\ref{vp0}) with the substitute
\begin{equation}
c_{s}^{2}+\frac{1}{2\ }V_{\phi }^{2}\rightarrow c_{s}^{2}+\frac{p_{\bot }}{%
nm_{i}\ }.
\end{equation}%
The solution of equation (\ref{vp0}) can also be written using the Lambert
function.

The profiles of the accelerating transonic solution for the plasma flow in
the mirror for different values of $M_{\phi 0}$ are shown in Fig.~3. These profiles were obtained using a magnetic nozzle with $R = 10$ and $K = 800$ for $z_0/L=0.5$ and for $B_\Vert$ in Eq. (40) in the interval $z_0<z<L$.  We note that in the MHD model and in neglect of the azimuthal magnetic field, Eqs. (51)--(53) are equivalent to the hydrodynamic model of the neutral compressible gas.

Note that the slope of the axial velocity is determined by the second derivative of the magnetic field profile [21]  $B_{\Vert }\left( z\right)$. 
Since the second derivative of the profiles in Eq. \eqref{Bm} is discontinuous at $z/L=0.5$, in Fig. 2 and other figures below, one can observe a break in the slope of $V_{\Vert }\left( z\right)$ at the point $z/L=0.5$. It is interesting to note the jump of the derivative of the ion flow observed in Figs. 3--7 is also reproduced in the first principle drift-kinetic simulations of the plasma flow in the magnetic nozzle [23].

\subsection{Effects of azimuthal magnetic field and azimuthal rotation} 

Equations (27)--(30) may create an impression that 
 the $V_\phi$ and $B_\phi$ variables are independent of the radial coordinate $r$.  As it is discussed in Section 5,  this is not so, since  the conservation  equations (27)--(30) are written for the magnetic flux tubes of a fixed value of the magnetic flux, and the dependence on the radial coordinate $r$ is implicit in Eqs. (27)--(30) via the dependence of the integral constants. 
The radial dependence can be made explicit by taking into account the  magnetic flux conservation condition $BS=B\pi r^{2}=\text{const}.$
 
 Using these condition, one can identify some familiar limits that follow from the integrals of motion given by Eqs. (\ref{bphi}) and (\ref{vphi}).
 In the
electrostatic limit, where plasma rotates with angular $\mathbf{E}\times
\mathbf{B}$ velocity, one obtains
\begin{equation}
{\ \ V_{\phi }\sim \ {-}\frac{\beta }{B_{\Vert }}\sqrt{B_{\Vert }}\sim -}%
\frac{\beta }{\sqrt{B_{\Vert }}}{\sim r.}  \label{vphia}
\end{equation}%
In the limit corresponding to pure hydrodynamics of a neutral fluid, one
has the rotation corresponding to the conservation of the angular momentum
\begin{equation}
\ V_{\phi }\sim \frac{W}{\Gamma }\sqrt{B_{\Vert }}\sim \frac{1}{r}.
\label{vphib}
\end{equation}%

Similarly,  for the azimuthal magnetic field one has two limits 
\begin{equation}
B_{\phi }\sim -\frac{W}{4\pi }\sqrt{B_{\Vert }}\sim \frac{1}{r}{,}
\end{equation}%
and
\begin{equation}
B_{\phi }\sim -\frac{\beta \Gamma }{{\sqrt{B_{\Vert }}}}\sim r,
\end{equation}%
respectively, characterizing the magnetic field outside of the linear
current, and the magnetic field near the $r=0$ axis.
 We note that with the replacement  $B_\Vert\rightarrow 2\psi /r^2$, equations (28), (29) become identical to the equations in the form used in Ref. [1], which explicitly contain radial variable $r$. 

 In the paraxial approximation, one can explicitly use the expressions  for the azimuthal rotation and magnetic field   near the axis $r=0$: $V_\phi=\Omega r$ and $B_\phi=J r$, where the angular rotation velocity and effective current density are the function of $z$, $\Omega=\Omega(z)$ and $J=J(z)$. Then, from equations (1)--(4) in the stationary case, one obtains, in the first order in $r$, the following equations: 
\begin{equation}
{V_{\parallel }J-\Omega B_{\parallel }=}{B_{\Vert }}\beta^\prime,
\end{equation}%
\begin{equation}
\Gamma \Omega-J/4\pi =B_{\Vert } W^\prime,  
\end{equation}where $\beta^\prime$ and $W^\prime $ are now the constant independent of the radial variable $r$. These equations are fully equivalent to equations (28), (29).
 
Electromagnetic effects are
important when plasma flow velocity reaches the Alfven and fast magnetosonic
velocity. At best, in the isothermal approximation, the  velocity of the
accelerated plasma increases with the expansion ratio of the magnetic field
logarithmically. Neglecting  the logarthmic  factor of the order of one, from mass
conservation, one  has for the plasma density  $\rho \sim B_{\Vert }.$ Then,
Alfven velocity scales as $c_{A}\sim $ $B_{\Vert }/\sqrt{4\pi \rho }\sim \sqrt{%
B_{\Vert }}.$ Thus, at high expansion,  plasma flow velocity will reach the
Alfven velocity, $V_{\Vert }\sim c_{A}$, and electromagnetic effects become
important. The azimuthal magnetic field and rotation are intrinsically
coupled in the ideal MHD model. This coupling is  somewhat similar to that in the circularly polarized Alfven wave, but here we consider  stationary, $\partial/\partial t =0 $, nonlinear structures  in the moving plasma and inhomogeneous axial magnetic field. Thus, the nonlinear solutions discussed here  involve axial flow $V_\Vert$, plasma pressure $p$, plasma rotation $V_\phi$, and azimuthal magnetic field $B_\phi$.  

Below we give several examples of such solutions for different boundary conditions at $z=0$.
As an example, Figs. 4 and 5 consider the
cases when either the azimuthal magnetic field or rotation velocity is
zero at the breach of the magnetic nozzle $z=0$.  In Fig. 4, with $%
B_{\phi }\left( 0\right) =0$ and $V_{\phi }\neq 0,$ one can observe the
generation of the magnetic field from the zero value $B_{\phi }\left(
0\right) =0$ at the breach. An alternative situation is shown in Fig.~5, with
$V_{\phi }=0$ and $B_{\phi }\left( 0\right) \neq 0,$ where one can see the
generation of the rotation along the nozzle driven by the azimuthal magnetic
field. Plasma acceleration by the combined effect of the finite $B_{\phi
}\left( 0\right) \neq 0$ and $V_{\phi }\left( 0\right) \neq 0$ is
illustrated in Fig.~6. The azimuthal rotation velocity $V_\phi$ was normalized to the ion-sound velocity, $M_\phi=V_\phi/c_s$.

For simplicity the calculations in Figs. 3--7 are
performed for the isothermal model and the magnetic field from Eqs. (39) and (40) was used with $R=10$ and $K=800$, with Eq. (40) used in the interval $z_0<z<L$, $z_0=0.5L$. For these parameters,  the slow magnetosonic, Alfven and fast magnetosonic critical points occur at $z/L = 0.5$, $z/L = 0.93$ and $z/L = 1.32$ respectively. Thus, the fast-magnetosonic point is outside the range shown in Figs.~3--7.

It is worth noting that the solutions with finite rotation $V_\phi$ and azimuthal magnetic field $B_\phi$,  shown in Figs. 4 and 5, involve both axial and radial currents, $J_z=r^{-1}\partial (r B_\phi)/\partial r  $ and $J_r=-\partial  B_\phi/\partial z$. Note, that as it was discussed above, $B_\phi=B_\phi (r,z)$, and the local current conservation, $\nabla \cdot \mathbf{J}=0$, is satisfied automatically. The situation considered in Fig.~4 has 
$B_{\phi }\left( 0\right) =0$ and thus it is globally ambipolar since the direction of the radial current changes with the sign of $B_\phi$ and the total radial is zero upon the integration over the whole volume.   The case shown in  Fig.~5 has 
$B_{\phi }\left( 0\right) \neq 0$ at $z=0$ and thus has a finite axial current $J_z$ entering the magnetic nozzle at $z=0$. Such situations are relevant  to the open magnetic systems with  biased electrodes that create the axial electric current [5, 6] and may also occur in hybrid electrostatic-magnetic hybrid acceleration schemes [7]. In this case, the net radial  current is finite and has to be closed at the outer boundary which can be achieved through the chamber walls [5, 6] or radial electrodes [7].  The closure problem is a design specific and is not considered here.

% \begin{figure}[h]
%     \centering
% \includegraphics[width=0.32\textwidth]{V7a.eps}
% \includegraphics[width=0.32\textwidth]{V7b.eps}
% \includegraphics[width=0.32\textwidth]{V7c.eps}
%     \caption{Axial profiles of $V_{\parallel}(z)$ (left),  azimuthal rotation $V_{\protect\phi}(z)$ (center), and azimuthal magnetic field $B_{\protect\phi }(z)$  (right) for $B_{\phi}(0) = 0.1$ and various values of $V_{\phi}(0)/c_{s}$. }
%     \label{V_phi_bc}
% \end{figure}

% \begin{figure}[h]
%     \centering
% \includegraphics[width=0.32\textwidth]{V9a.eps}
% \includegraphics[width=0.32\textwidth]{V9b.eps}
% \includegraphics[width=0.32\textwidth]{V9c.eps}
%     \caption{Axial profiles of $V_{\parallel}(z)$ (left),  azimuthal rotation $V_{\protect\phi}(z)$ (center), and azimuthal magnetic field $B_{\protect\phi }(z)$  (right) for different Dirichlet boundary conditions  at the left end and free boundary conditions at the right end for $V_{\parallel}$. }
%     \label{V_phi_bc}
% \end{figure}

\section{Time dependent dynamics and stability of stationary transonic and trans-Alfvenic solutions}

As an alternative to stationary solutions obtained with equations \eqref{s1}--\eqref{sg}, as for the solution diagram in Fig. 2, we also have studied the time-dependent evolution in the initial value simulations using equations (15)--(18). These nonlinear equations were solved numerically using BOUT++ solver \red{[24]} on a grid of 252 points in the spatial domain and applying the appropriate boundary and initial conditions for each variable. In the simulations $V_{\parallel}$ and $V_{\phi}$, as shown in Figs. 4--7, were normalized using the speed of sound $c_{s} = \sqrt{T_{e}/m_{i}}$ such that $M_{\parallel} = V_{\parallel}/c_{s}$, $M_{\phi} = V_{\phi}/c_{s}$. The plasma density and magnetic field were normalized to the values at the left end of the nozzle, $z/L = 0$,  $n' = n/n_{0} = \rho/\rho_{0}$,  $B'_{\phi} = B_{\phi}/B_{\parallel 0}$ and $B'_{\parallel} = B_{\parallel}/B_{\parallel 0}$. For our parameters, the characteristic time $t_{c} = L/c_{s} = 8.3 \times 10^{-6}$~s.  We applied Dirichlet boundary conditions for $n/n_0$, $V_{\phi}/c_{s}$, and $B_{\phi}/B_{\parallel 0}$ at $z/L = 0$ and free boundary conditions at $z/L = 1$. The variable $V_{\parallel}/c_{s}$ had free boundary conditions at both ends. The Dirichlet boundary conditions for each variable are shown in the legend of the appropriate figure. Each variable had a constant value throughout the entire nozzle, equal to the value at the $z/L = 0$ boundary, as an initial condition. As an example, the case in Fig. 6 represented by the red dashed curve had the following left end boundary and initial conditions: $n(z/L= 0)/n_{0} = 1.0$, $n(t = 0)/n_{0} = 1.0$, $V_{\phi}(z/L = 0)/c_{s} = 0.5$, $V_{\phi}(t = 0)/c_{s} = 0.5$, $B_{\phi}(z/L)/B_{\parallel 0} = 0.1$, $B_{\phi}(t = 0)/B_{\parallel 0} = 0.1$, and each variable had free boundary conditions at the right end of the nozzle, $z/L = 1$. The initial value for $V_{\parallel}/c_{s}$ was 0.1 in normalized units. The simulation was run for 10000 time-steps with each time-step $\delta t=0.001t_{c}$. The simulation was terminated once the stationary solutions were well established, typically around a few $t_c$. The profiles of  $n$, $V_{\parallel}$, $V_{\phi}$ and $B_{\phi}$ obtained in steady state, were used in equation (27)--(30) to confirm that the values of $\Gamma$, $\beta$, $W$ and $I$ remain constant throughout the entire nozzle.

 These simulations show
that a wide range of initial states converge in time to the profiles obtained
from stationary equations. An example of such evolution is shown in Fig.~\ref{time-evolution}, where the solution with a free boundary condition for $V_\Vert$ at $z=0$ converges to a unique transonic solution. We note that not every arbitrary choice of values for $V_{\phi}$ and $B_{\phi}$ at $z=0$ result in a valid transonic and trans-Alfvenic solution, as it is seen in Fig. 2. Modification of the boundary values for $V_\Vert$,  $V_{\phi}$, and $B_{\phi}$ result in distortion of the diagram of stationary solutions in Fig, 2. But some boundary values produce unphysical solutions in Fig. 2, e. g. the physically inadmissible multiple value solutions. The robustness of the unique transonic and trans-Alfvenic solution is seen in the behavior of the stationary solutions in the time-dependent initial value simulations in Figs. 4--7. One can see that for some sets of rather arbitrary chosen values of $V_{\phi}$, and $B_{\phi}$  at $z=0$, there is a narrow boundary (transition) layer at $z=0$.  In the result, the system ``finds'' the ``correct'' boundary values for these plasma parameters and converges toward such a unique solution. Such behavior has also been seen in other fluid and kinetic simulations \red{[14, 25]}.

% It was found that $\Gamma$, $\beta$, $W$ and $I$ were indeed constant throughout the entire nozzle and theses constants were used in equations 37 - 38 to obtain values for $V_{\phi}$ and $B_{\phi}$. The values of $V_{\phi}$ and $B_{\phi}$ obtained using the constants were then compared to the values of said variable obtained from the stationary solution and it was found that both sets of values were indeed equal and produced the same axial profile. The values of the constants were used together with the values of $V_{\parallel}$ and $B_{\parallel}$ in the Bernoulli integral to obtain the solution curves of Fig. 1.

\section{Conclusion}

We have shown here that within a generalized electromagnetic Laval nozzle problem,
the energy of the magnetic field and plasma rotation can be converted into
the kinetic energy of the directed (accelerated) plasma flow, similar to
the conversion of thermal energy in a gas-dynamics Laval nozzle. An
important
property of plasma flow through the nozzle is that it follows a unique
profile fully defined by the magnetic field, i. e. the transonic
accelerating solution has no free parameters. The robust nature of the transonic accelerating solution was shown in the fluid solutions of Refs. \red{[14, 19]} and further confirmed in
kinetic numerical simulations \red{[23, 25]}. We note that, in realistic applications, the constraint in the transonic solution on the plasma flow velocity at the breach of the magnetic nozzle translates into an adjustment of the plasma density in the source region \red{[23, 25]}.
In other words, the plasma
source is matched with the transonic solution at the breach of the magnetic nozzle via the generation of an electric field and partial particle reflections by the electric field and mirror force, as shown in Ref. {[25]}.

The unique transonic and trans-Alfvenic accelerating solution is constrained
by the regularization requirements at all three critical points: slow
magnetosonic, Alfven, and fast magnetosonic, where the flow velocity is
equal to the phase velocity of the respective mode. The three critical
points define the topology of the possible solutions on the ``phase-space'' $%
\left( V_{\Vert }\left( z\right) ,z\right) $ diagram, Fig. 2. The solutions
at critical points are regularized by the combinations of the initial
conditions and $\partial B/\partial z=0$ condition at the magnetic nozzle
throat.

It is shown here that the magnetic field energy $B_{\phi }^{2}/8\pi $ and
the azimuthal rotation energy $\rho V_{\phi }^{2}/2$ can be
effectively converted into the kinetic energy of the accelerated flow. These
properties of the magnetic nozzle can be useful in maximizing the thrust
density in the intermediate regimes between the classical annular Hall
thruster configurations and the diverging magnetic field (magnetic nozzle)
configurations, such as in the cylindrical Hall thruster and related
wall-less configurations [7, 26, 27]. Potentially, such regimes will be able to combine
the high performance of annular Hall thrusters [28] with the additional
thrust and flexibility provided by thrusters with a diverging magnetic
field (magnetic nozzle) and magnetically driven (MPD type) thrusters.

This paraxial model does not address the diamagnetic effects caused by
azimuthal currents [29, 30], which require a two-fluid [31] and fully two-dimensional
theory. The paraxial model considered here assumes a constant poloidal
magnetic field profile $B_{\Vert }\left( z\right).$ The azimuthal currents
result in modifications of the poloidal magnetic field $B_{\Vert }\left(
z\right),$ which is most likely to occur at the fast-magnetosonic critical
point. At this point, the plasma flow may tear the external magnetic field
via reconnection due to small resistivity (collisional or turbulent); the
plasma flow will detach, carrying the magnetic flux away with the flow [32].  For a finite density of a neutral gas, charge-exchange reactions may also provide dissipation affecting plasma acceleration [14], reconnection and subsequent plasma detachment. Such effects have to be included together with the electromagnetic effects studied in this paper. The electromagnetic effects are central to the plasma acceleration in magnetoplasma compressors [33,34] and may also be important for supersonic jets of dense plasmas [35].

\section*{Acknowledgments}

This research is supported by Natural Sciences and Engineering Research
Council of Canada, and the University of Saskatchewan.

\newpage

\section*{FIGURES CAPTIONS}

\textbf{Fig. 1.} Magnetic flux tube formed by an inhomogenous magnetic field $\mathbf{B}$ of the nozzle.

\textbf{Fig. 2.} The ``phase-space'' diagram of multiple solutions of stationary equations (\ref{s1})--(\ref{sg}) demonstrating electromagnetic effects for large expansion. The slow magnetosonic (former sonic point) critical point is around $z/L=0.5$, i. e. at the magnetic nozzle throat. The Alfven and fast magnetosonic critical points are near $z/L\approx 4.7$ and $z/L\approx 8.5$ respectively. At  these  locations the mirror ratios are $1$, $640$ and $2270$, respectively.

\textbf{Fig. 3.} Swirl acceleration: axial profiles of $V_{\parallel}$ \red{(a)} and azimuthal rotation $V_{\protect\phi}$ \red{(b)}. Azimuthal magnetic field is neglected, $B_{\protect\phi }\equiv 0.$  Magnetic field from Eq. (39) and Eq. (40) in the interval $z_0<z<L$, $z_0=0.5L$ were used.

\textbf{Fig. 4.} Plasma flow  $V_{\parallel}(z)$ \red{(a)} and azimuthal rotation $V_{\phi}(z)$ \red{(b)} for $V_{\protect\phi}(0) \ne 0 $,  $B_{\protect\phi }(0)=0$. Note generation of $B_{\protect\phi}(z)$ \red{(c)} along the magnetic nozzle. 

\textbf{Fig. 5.} Plasma flow  $V_{\parallel}(z)$ \red{(a)} and azimuthal magnetic field $B_{\phi}(z)$ \red{(b)} for $B_{\protect\phi }(0) \ne 0$ and $V_{\protect\phi}(0) = 0 $. Note generation of azimuthal rotation $V_{\protect\phi}(z)$ \red{(c)} along the magnetic nozzle.

\textbf{Fig. 6.} Axial profiles of $V_{\parallel}(z)$ \red{(a)},  azimuthal rotation $V_{\protect\phi}(z)$ \red{(b)}, and azimuthal magnetic field $B_{\protect\phi }(z)$  \red{(c)} for $V_{\protect\phi}(0)\ne 0 $ and $B_{\protect\phi}(0)\ne 0$.

\textbf{Fig. 7.} Axial profiles of $V_{\parallel}(z)$ \red{(a)},  azimuthal rotation $V_{\protect\phi}(z)$ \red{(b)}, and azimuthal magnetic field $B_{\protect\phi }(z)$  \red{(c)} for $V_{\phi}(0)/c_{s} = 0.5$ and various values of $B_{\phi}(0)$.

\textbf{Fig. 8.} Time evolution of the plasma velocity of $V_{\parallel}(z,t)$ \red{(a)} and plasma density $n(z,t)$ \red{(b)} toward the stationary equilibrium solution as described in Section~7.

\cleardoublepage 
\begin{figure}[h]
	\centering
	\includegraphics[width=0.60\textwidth]{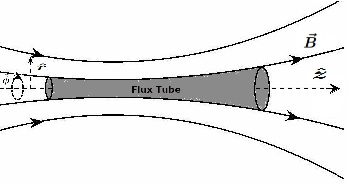}
	\caption{ } 
	\label{magnetic_nozzle}
\end{figure}

\begin{figure}[h]
	\centering
	\includegraphics[width=0.4\textwidth]{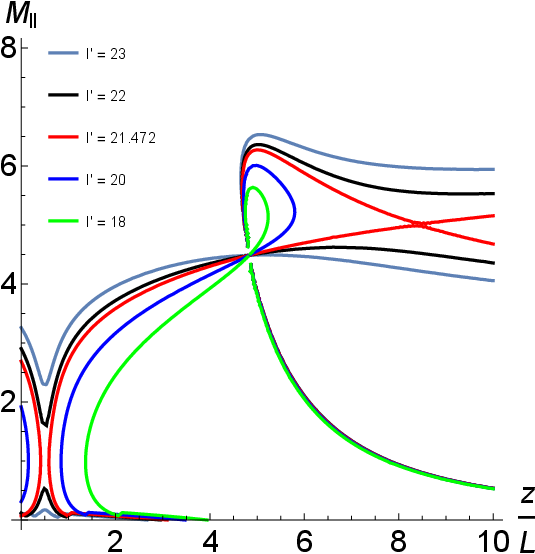}
	\caption{ }
	\label{fig:sidebyside}
\end{figure}

\begin{figure}[h]
	\centering
	\includegraphics[width=0.4\textwidth]{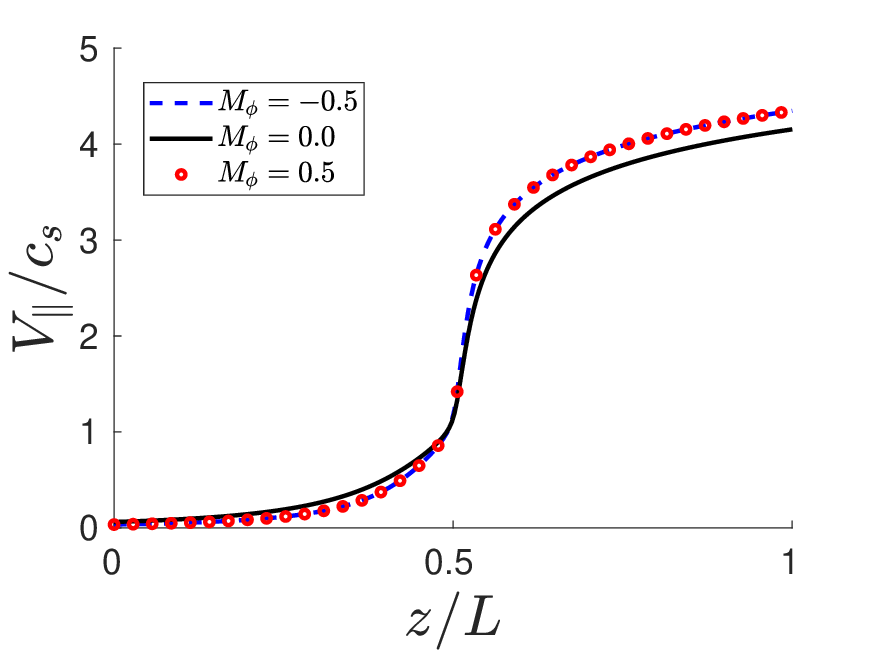}
	\includegraphics[width=0.4\textwidth]{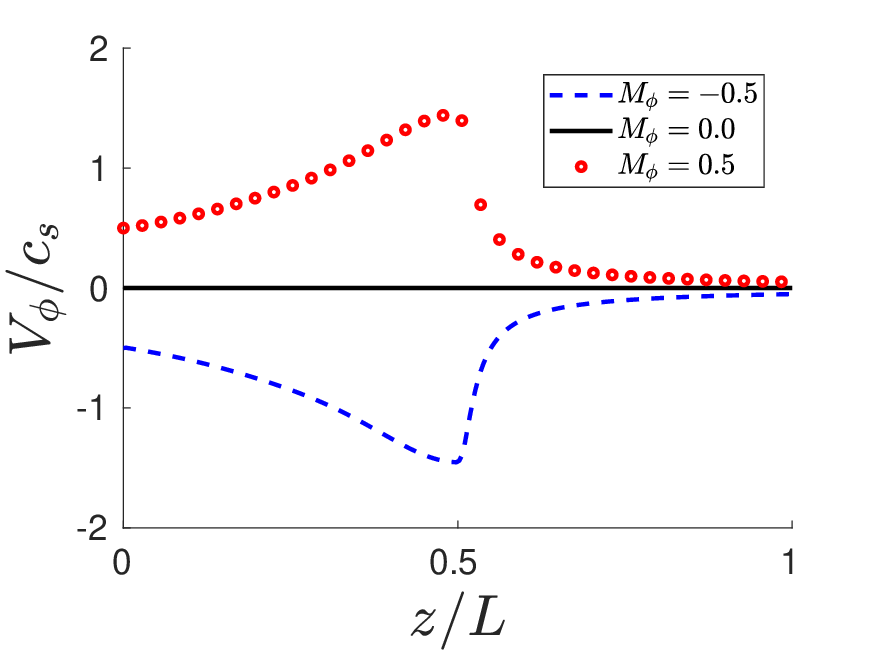}\\
	a\qquad\qquad\qquad\qquad\qquad\qquad b
	\caption{ }
	\label{fig:sidebyside}
\end{figure}

\begin{figure}[h]
	\centering
	\includegraphics[width=0.32\textwidth]{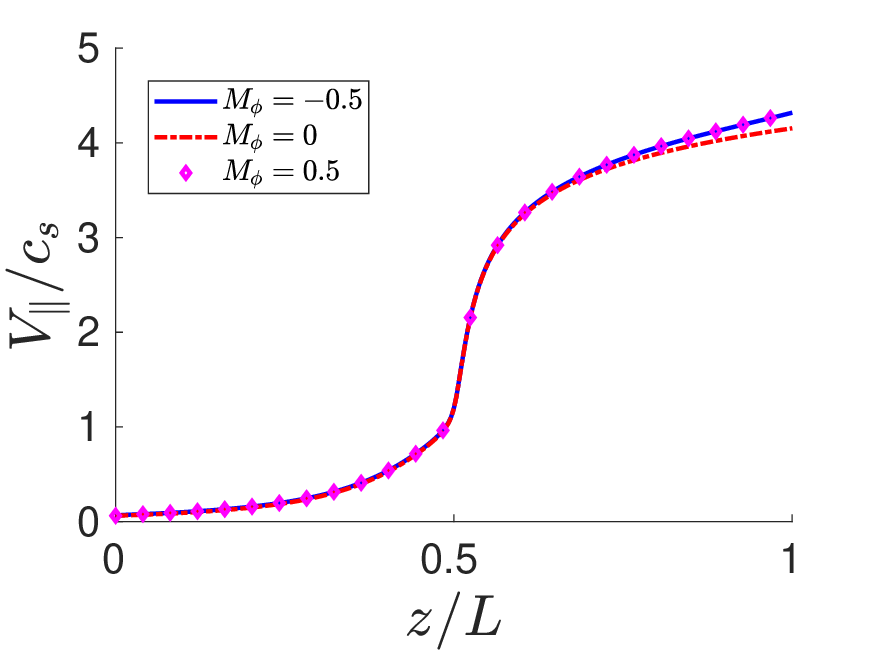}
	\includegraphics[width=0.32\textwidth]{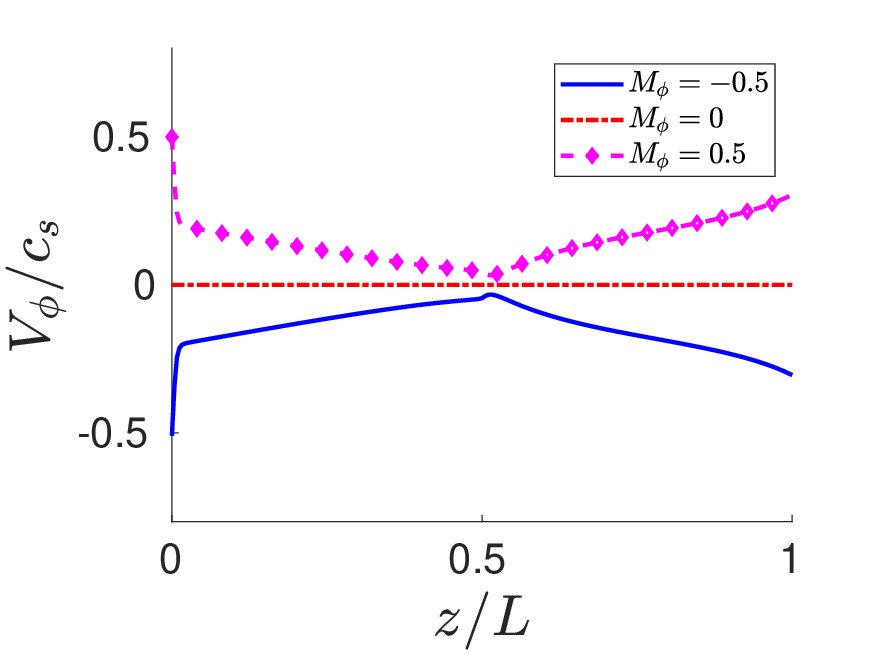}
	\includegraphics[width=0.32\textwidth]{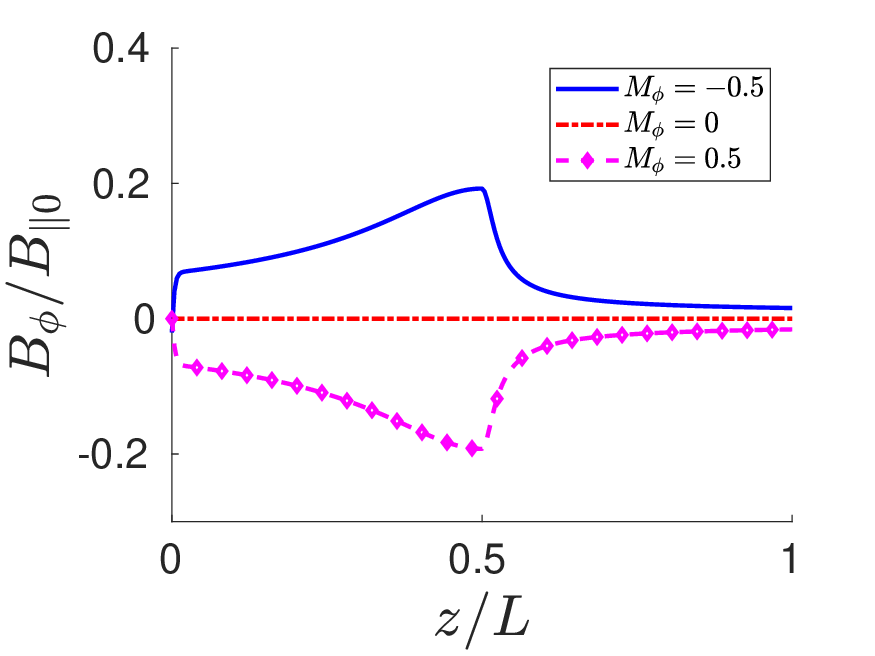}\\
	a\qquad\qquad\qquad\qquad\qquad\qquad b\qquad\qquad\qquad\qquad\qquad\qquad c
	\caption{ }
	\label{fig:sidebyside}
\end{figure}

\begin{figure}[h]
	\centering
	\includegraphics[width=0.32\textwidth]{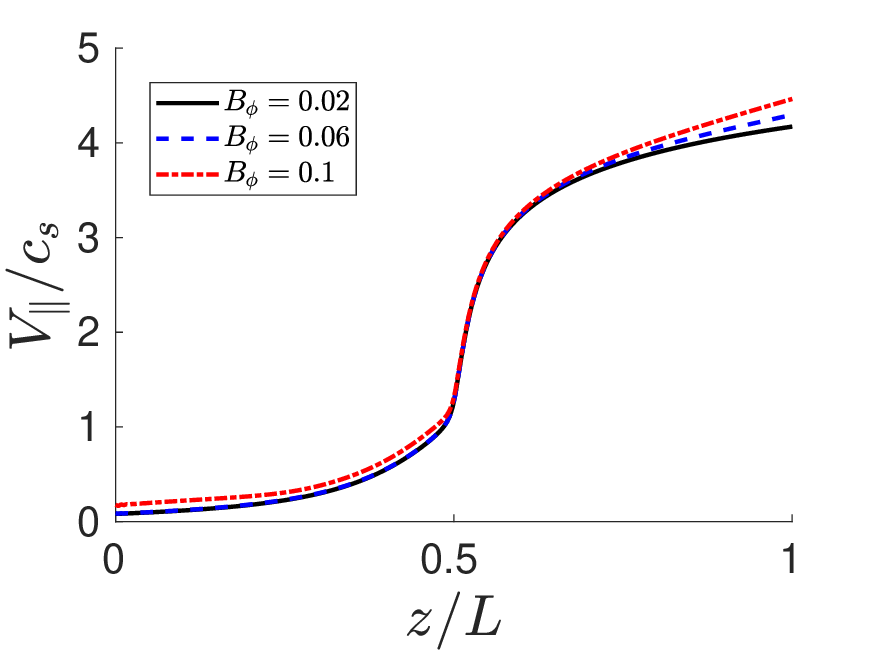}
	\includegraphics[width=0.32\textwidth]{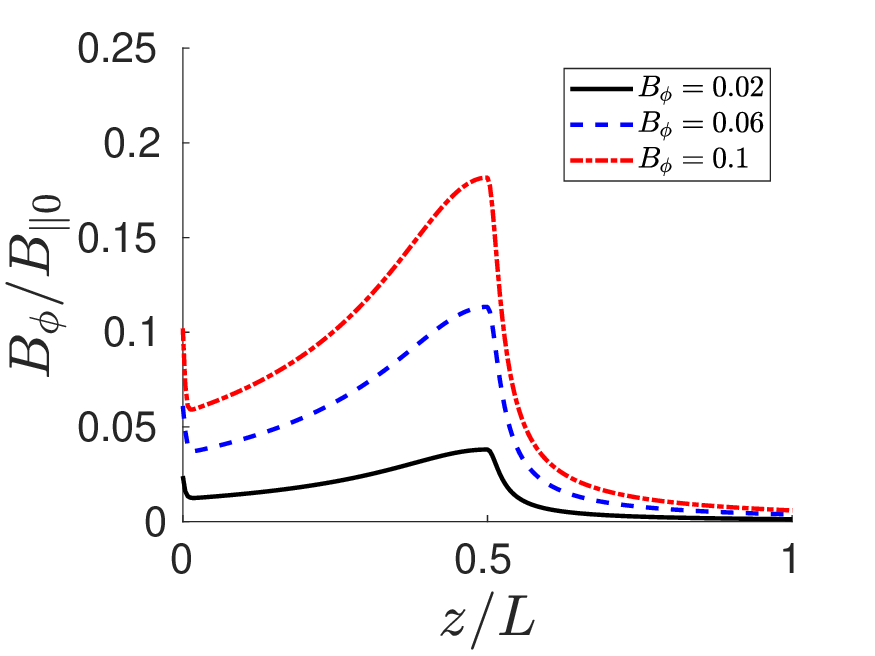}
	\includegraphics[width=0.32\textwidth]{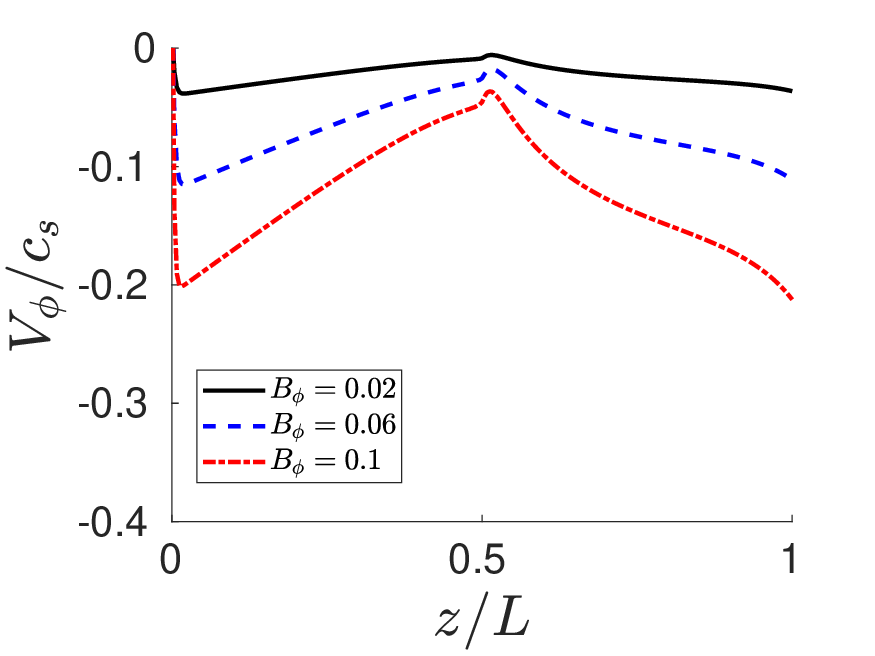}\\
	a\qquad\qquad\qquad\qquad\qquad\qquad b\qquad\qquad\qquad\qquad\qquad\qquad c
	\caption{ }
	\label{fig:3}
\end{figure}

\begin{figure}[h]
	\centering
	\includegraphics[width=0.32\textwidth]{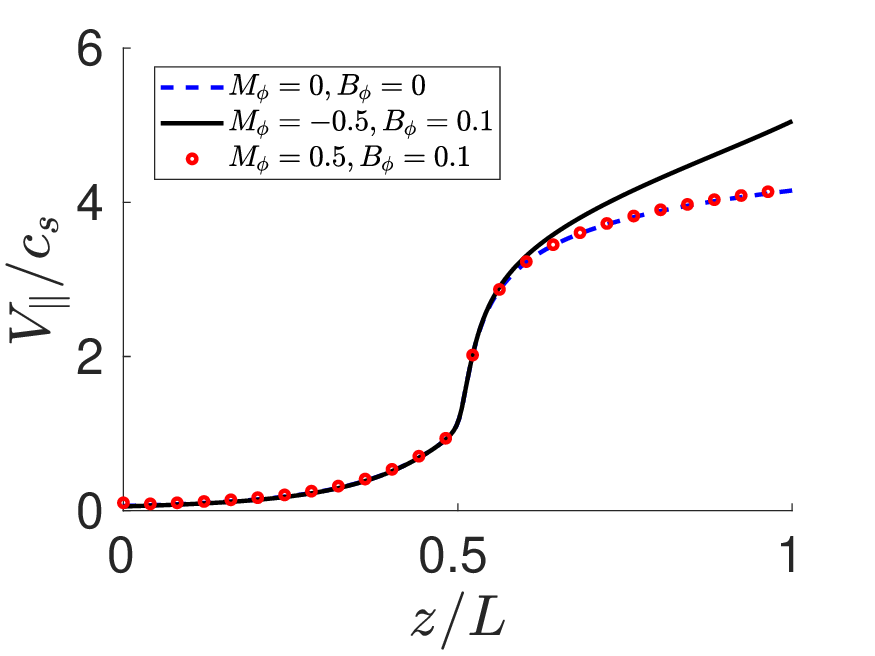}
	\includegraphics[width=0.32\textwidth]{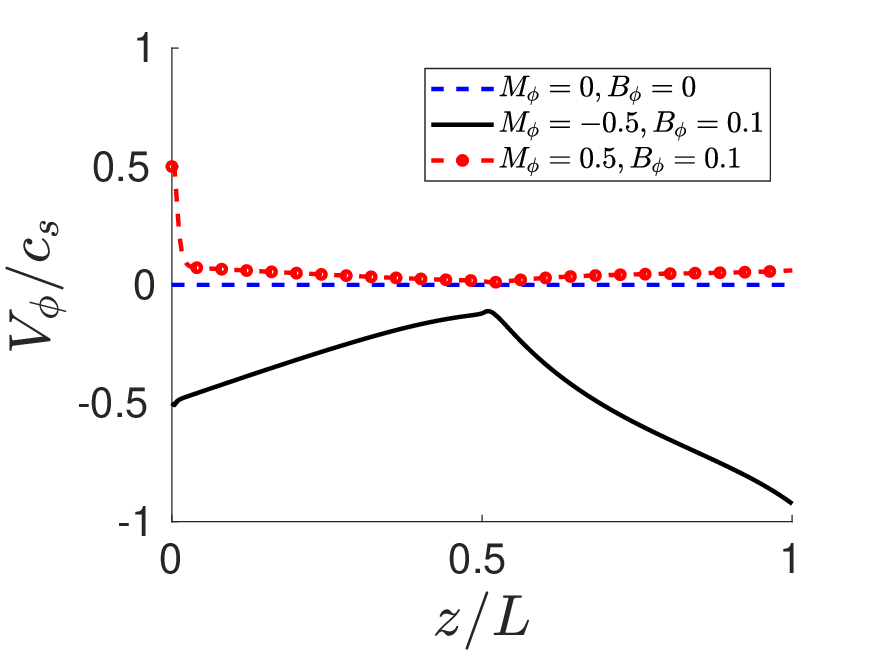}
	\includegraphics[width=0.32\textwidth]{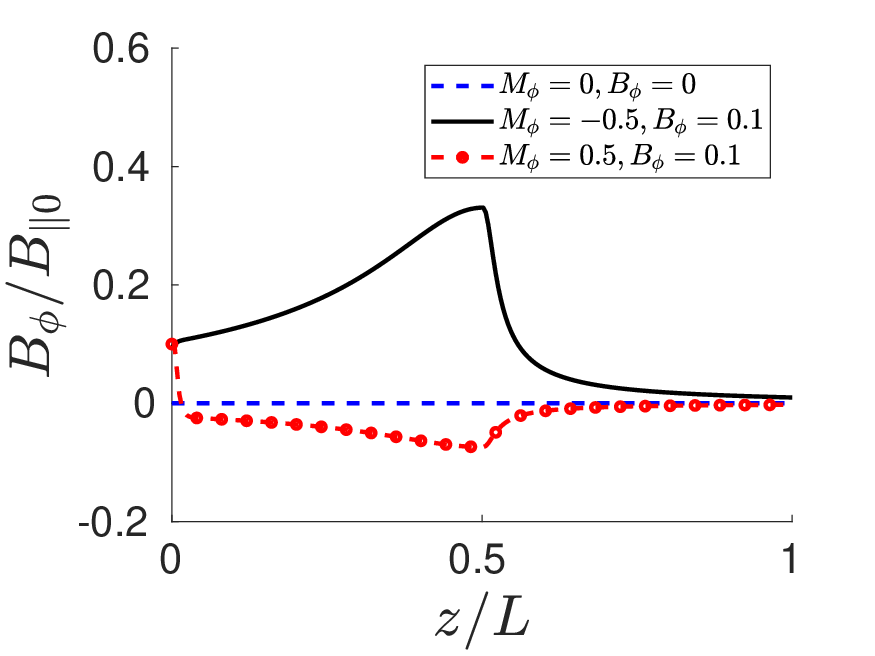}\\
	a\qquad\qquad\qquad\qquad\qquad\qquad b\qquad\qquad\qquad\qquad\qquad\qquad c
	\caption{ }
	\label{fig:3}
\end{figure}

\begin{figure}[h]
	\centering
	\includegraphics[width=0.32\textwidth]{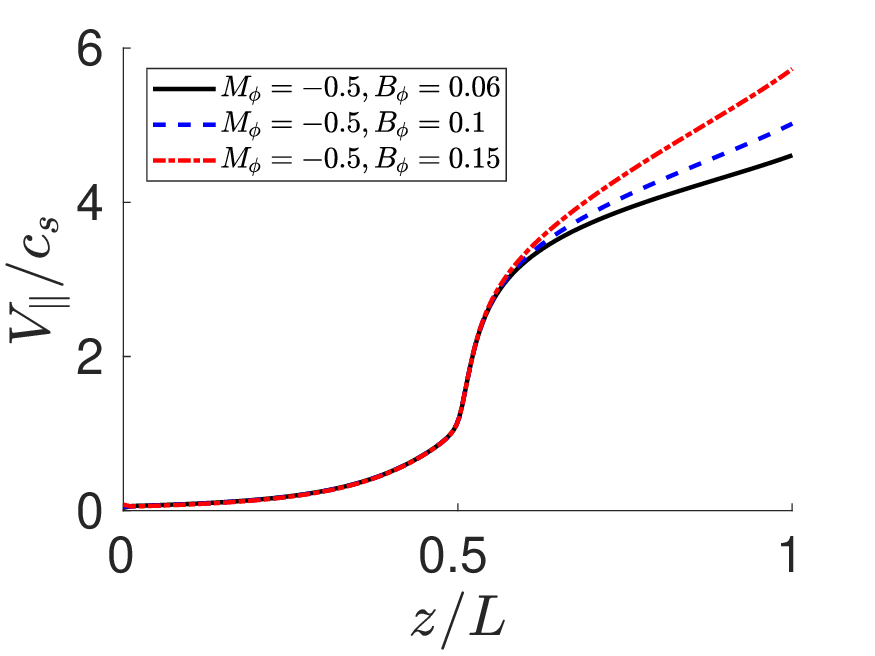}
	\includegraphics[width=0.32\textwidth]{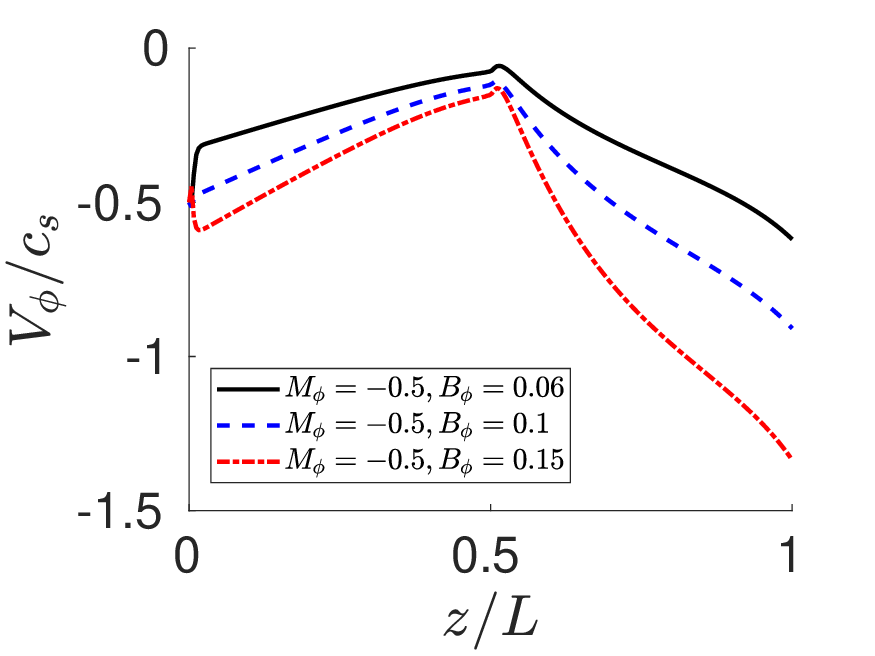}
	\includegraphics[width=0.32\textwidth]{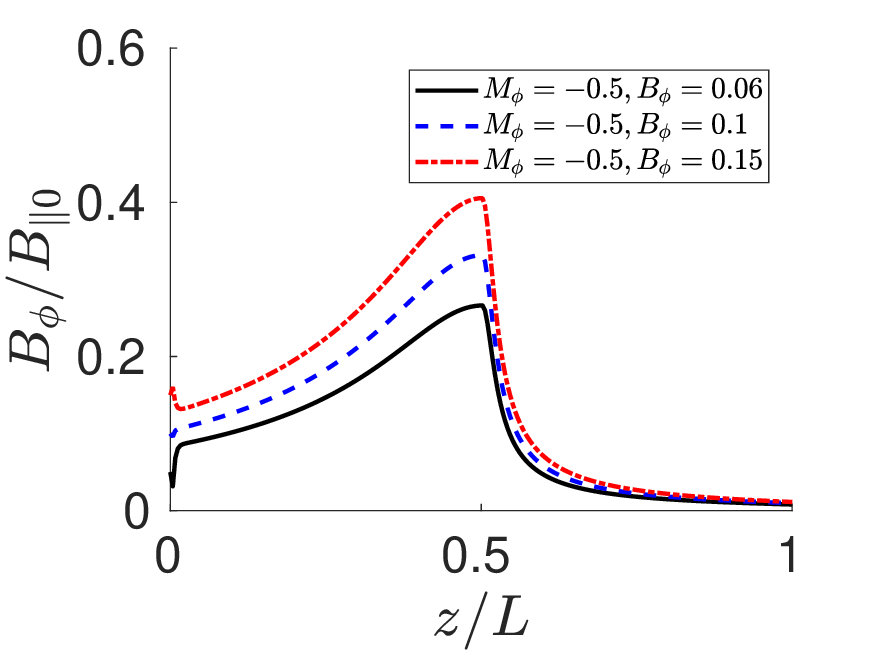}\\
	a\qquad\qquad\qquad\qquad\qquad\qquad b\qquad\qquad\qquad\qquad\qquad\qquad c
	\caption{ }
	\label{B_phi_bc}
\end{figure}
\begin{figure}[h]
	\centering
	\includegraphics[width=0.32\textwidth]{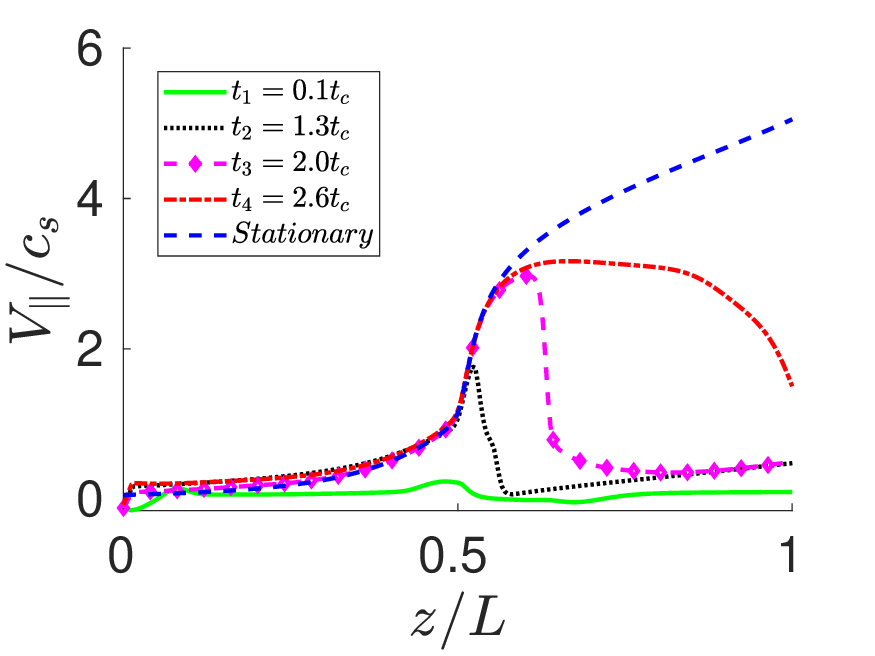}
	\includegraphics[width=0.32\textwidth]{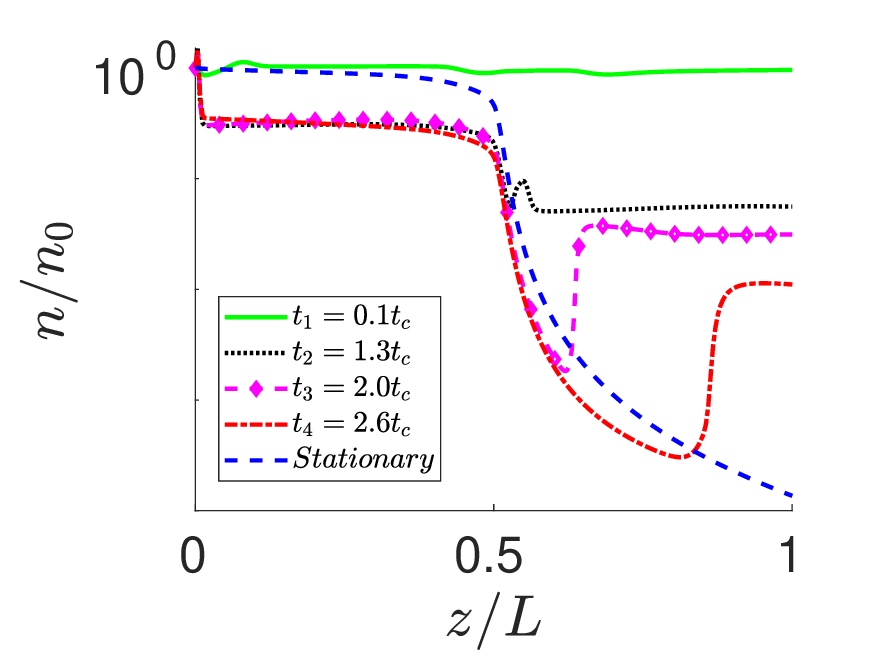}\\
	a\qquad\qquad\qquad\qquad\qquad\qquad b
	\caption{ }
	\label{time-evolution}
\end{figure}

\end{document}